\documentclass[english]{article}
\usepackage[T1]{fontenc}
\usepackage[latin9]{luainputenc}
\usepackage{geometry}
\usepackage{amstext}
\usepackage{amssymb}
\usepackage{feyn}
\usepackage{tikz}
\usepackage{graphicx}
\usepackage{array}
\usepackage{booktabs}

\usepackage{amsmath}
\usepackage{amsfonts}

\usepackage[normalem]{ulem}
\usepackage{color}
\usepackage{listings,braket}
\usepackage{caption}
\usepackage{subcaption}
\usepackage{float}
\definecolor{darkgreen}{rgb}{0,0.35,0}
\definecolor{Rood}{rgb}{1, 0, 0}

\newcommand{\hvarphi}{\hat \varphi}

\makeatletter
\usepackage[english]{babel}
\usepackage{feyn}

\makeatother

\begin{document}

\title{\textbf{Remarks on the Clauser-Horne-Shimony-Holt inequality in relativistic quantum field theory}}

{\author{\textbf{G.~Peruzzo$^1$}\thanks{gperuzzofisica@gmail.com},
\textbf{S.~P.~Sorella$^1$}\thanks{silvio.sorella@gmail.com},\\\\\
\textit{{\small $^1$UERJ -- Universidade do Estado do Rio de Janeiro,}}\\
\textit{{\small Instituto de F\'{\i}sica -- Departamento de F\'{\i}sica Te\'orica -- Rua S\~ao Francisco Xavier 524,}}\\
\textit{{\small 20550-013, Maracan\~a, Rio de Janeiro, Brasil}}\\
}

\date{}

\maketitle
\begin{abstract}
We present an investigation of the {\it CHSH} inequality within a relativistic quantum field theory model built up with a pair of free massive scalar fields $(\varphi_A, \varphi_B)$ where, as it is customary,  the indices $(A,B)$  refer to Alice and Bob, respectively. A set of bounded Hermitian operators is introduced by making use of the Weyl operators. A {\it CHSH} type correlator is constructed and evaluated in the Fock vacuum by means of the canonical quantization. Although the observed violation of the {\it CHSH} inequality turns out to be rather small as compared to Tsirelson's bound of Quantum Mechanics, the model can be employed for the study of Bell's inequalities in the more physical case of gauge theories such as: the Higgs models, for which local $BRST$ invariant operators describing both the massive gauge boson as well as the Higgs particle have been devised. These operators can be naturally exponentiated, leading to $BRST$ invariant type of Weyl operators useful to analyze Bell's inequalities within an invariant $BRST$ environment.  
\end{abstract}

\section{Introduction} 
Since their discovery \cite{Bell:1964kc,Bell:1964fg,Bell:1980wg,Bell:1987hh}, Bell's inequalities have much changed the way we look at the quantum world, forcing us to go deeper and deeper in the understanding of Quantum Mechanics and of the nature of space-time. It is fair to say that, nowadays, the phenomenon of entanglement  is a pivotal issue in both theoretical and experimental physics as well as in the creation of new technologies. \\\\This work aims at investigating, within the framework of relativistic Quantum Field Theory,  a very popular and extensively studied version of Bell's inequalities, known as the $CHSH$ inequality \cite{Clauser:1969ny,Freedman:1972zza,Clauser:1974tg,Clauser:1978ng}. Let us briefly remind it, in the form usually presented in Quantum Mechanics textbooks, see for example \cite{Nielsen,Peres,Zwiebach}. One starts by introducing a two spin $1/2$ operator
\begin{equation}
{\cal C}_{CHSH} = \left[ \left( {\vec \alpha} \cdot {\vec\sigma}_A    + {\vec \alpha'} \cdot {\vec\sigma}_A  \right) \otimes  {\vec \beta}\cdot{\vec \sigma}_B + \left( {\vec \alpha} \cdot {\vec \sigma}_A    - {\vec \alpha'} \cdot {\vec\sigma}_A  \right) \otimes  {\vec \beta'}\cdot{\vec \sigma}_B \right] \;, \label{chshqm}
\end{equation}
where $(A,B)$ refer to Alice and Bob, $\vec \sigma$ are the spin $1/2$ Pauli matrices and $({\vec \alpha},{\vec \alpha'},{\vec \beta},{\vec \beta'})$ are four arbitrary unit vectors.\footnote{Notice that due to $\sigma_i \sigma_j = \delta_{ij} + i\varepsilon_{ijk}\sigma_k$, it follows that $({\vec n} \cdot {\vec\sigma})^2 =1$ for any unit vector $|\vec n|=1$. }
Due to the properties of the Pauli matrices, one expects that 
\begin{equation} 
| {\cal C}_{CHSH} | \le 2 \;, \label{in}
\end{equation} 
for any possible choice of the unit vectors $({\vec \alpha},{\vec \alpha'},{\vec \beta},{\vec \beta'})$. Though, it turns out that this inequality is  violated by Quantum Mechanics, due to entanglement. In fact, when evaluating the $CHSH$ correlator in Quantum Mechanics, {\it i.e.} $ \langle \psi | {\cal C}_{CHSH} |\psi \rangle $, where $|\psi\rangle$ is an entangled state as, for example, the Bell singlet, one gets 
\begin{equation}
| \langle \psi | {\cal C}_{CHSH} |\psi \rangle | = 2\sqrt{2} \;, \qquad |\psi\rangle = \frac{ |+\rangle_A\otimes |- \rangle_B - 
|-\rangle_A\otimes |+\rangle_B}{\sqrt{2} }  \label{vb} \;. 
\end{equation} 
The bound $2\sqrt{2}$ is known as Tsirelson's bound \cite{tsi1,tsi2,tsi3}, yielding the maximum violation of the $CHSH$ inequality \eqref{in}. The experiments carried out over the last decades, see \cite{Clauser:1969ny,Freedman:1972zza,Clauser:1974tg,Clauser:1978ng,aspect0,aspect1,aspect2,aspect3,z0,z1} and refs therein, have largely confirmed the violation of the $CHSH$ inequality, being in very good agreement with the bound $2\sqrt{2}$. \\\\Concerning now the status of the study of the $CHSH$ inequality  within the relativistic Quantum Field Theory framework, the amount of research done so far cannot yet be compared to that of Quantum Mechanics. We quote here the pioneering work by \cite{Summers:1987fn,Summ,Summers:1987ze,Summers:1988ux,Summers:1995mv} who have been able to show, by using the techniques of the Algebraic Quantum Field Theory, that even free fields lead to a violation of the $CHSH$ inequality. This important result is taken as a strong confirmation of the fact  that the phenomenon of entanglement in Quantum Field Theory is believed to be  more severe than in Quantum Mechanics, a property often underlined in the extensive literature  on the so-called entanglement entropy, a fundamental quantity in order to quantify the degree of entanglement of a very large class of systems, see \cite{Harlow,Witten:2018zxz,Nishioka:2018khk,Casini:2022rlv} for recent overview on this matter. \\\\It seems thus worth to us to pursue the investigation of the $CHSH$ inequality within the realm of relativistic Quantum Field Theory. \\\\The paper is organized as follows. In Section \eqref{class}, we present the classical aspects of our field theory model as well as the class of  operators eligible in order to construct the $CHSH$ inequality. In Section \eqref{canon} we proceed with the canonical quantization and with the evaluation of the correlator of the $CHSH$ operator. Although rather small,  we shall be able to already observe a violation of the $CHSH$ inequality, confirming in fact the severity of entanglement in relativistic Quantum Field Theory.  In Section \eqref{CHSH}, the violation of the $CHSH$ is analyzed in details. Section \eqref{Hg} deals with the $BRST$ invariant generalization of the present setup to Higgs gauge theories.  

\section{The model: classical aspects} \label{class}
As already stated, the model we shall be using is constructed with a pair of free massive real scalar fields $(\varphi_A^i, \varphi_B^i)$, $i=1,2,3$, taken  in the adjoint representation of the $SU(2)$ group: 
\begin{equation} 
{\cal L} =   \frac{1}{2} \left( \partial^\mu \varphi_A^i \partial_\mu \varphi_A^i - m^2_A \varphi_ A^i\varphi_A^i \right)+\frac{1}{2} \left( \partial^\mu \varphi_B^i \partial_\mu \varphi_B^i - m^2_B \varphi_B^i\varphi_B^i \right) \;. \label{clact}
\end{equation} 
Further, we introduce the following bounded operator:
\begin{equation} 
{\cal{U}}^a(x,y) = \cos a^i(\hvarphi_A^i(x) + \hvarphi_B^i(y) )  = \frac{ e^{ia^i(\hvarphi_A^i(x) + \hvarphi_B^i(y) )} + e^{-ia^i(\hvarphi_A^i(x) + \hvarphi_B^i(y) )}}{2}\;, \label{Ua}
\end{equation} 
where $\{a^i\}$ stands for an arbitrary real vector and where we have introduced the rescaled fields $(\hvarphi_A^i,\hvarphi_B^i)$ in order to deal with dimensionless variables 
\begin{equation} 
 \hvarphi_A^i =\frac{\hvarphi_A^i}{m_A} \;, \qquad
\hvarphi_B^i=\frac{\hvarphi_B^i}{m_B} \;. \label{rescaled} 
\end{equation}
As it is apparent from expression \eqref{Ua}, the quantity ${\cal{U}}^a(x,y)$ is real and bounded, taking values in the interval $[-1,1]$. As such, according to \cite{Summers:1987fn,Summ,Summers:1987ze,Summers:1988ux,Summers:1995mv}, it is an eligible operator for the construction of a $CHSH$ inequality which, using the same notations of \cite{Summers:1987fn,Summ,Summers:1987ze,Summers:1988ux,Summers:1995mv}, we write as 
\begin{equation}
(A+A')B + (A-A')B' \;, \label{AB}
\end{equation}
with $(A,A',B,B')$ bounded quantities which take values in the interval $[-1,1]$. Application of the triangle inequality \cite{tsi1,tsi2,tsi3} shows that 
\begin{equation}
\bigl| (A+A')B + (A-A')B' \bigl| \le 2 \;. \label{AB2}
\end{equation}
In terms of the operator $\cal U$, expression \eqref{AB} takes the form  
\begin{equation}
{\cal C}^{aa'bb'}(x,x',y,y')  =  \left( {\cal{U}}^{a}(x,y)+{\cal{U}}^{a'}(x,y)\right) {\cal{U}}^{b}(x',y') + \left( {\cal{U}}^{a}(x,y)-{\cal{U}}^{a'}(x,y)\right) {\cal{U}}^{b'}(x',y') \;, \label{uu}
\end{equation} 
namely 
\begin{eqnarray}
{\cal C}^{aa'bb'}(x,x',y,y') & = &  \left[ \cos a^i (\hvarphi_A^i(x) + \hvarphi_B^i(y) ) + \cos a'^i(\hvarphi_A^i(x) + \hvarphi_B^i(y))  \right] \cos b^i (\hvarphi_A^i(x') + \hvarphi_B^i(y') )  \nonumber \\
& + & \left[ \cos a^i (\hvarphi_A^i(x) + \hvarphi_B^i(y) ) - \cos a'^i (\hvarphi_A^i(x) + \hvarphi_B^i(y) ) \right]\cos b'^i(\hvarphi_A^i(x') + \hvarphi_B^i(y') )  \;, \label{cos}
\end{eqnarray} 
with $(a^i,a'^i,b^i,b'^i)$ being arbitrary vectors. These vectors are akin to the four unit vectors  $({\vec \alpha},{\vec \alpha'},{\vec \beta},{\vec \beta'})$ entering expression  \eqref{chshqm}. Though, unlike $({\vec \alpha},{\vec \alpha'},{\vec \beta},{\vec \beta'})$, $(a^i,a'^i,b^i,b'^i)$ are now not restricted to be unit vectors, due to the fact that expression \eqref{Ua} is already bounded, taking values in the interval $[-1,1]$. They are independent quantities which, as the four vectors  $({\vec \alpha},{\vec \alpha'},{\vec \beta},{\vec \beta'})$, will be chosen in the most convenient way at the end of the computation. From equation \eqref{AB2}, we have that, classically, 
\begin{equation}
\l| {\cal C}^{aa'bb'}(x,x',y,y') \l| \;\le 2 \;. \label{le2}
\end{equation}
for any choice of the vectors $(a^i,a'^i,b^i,b'^i)$. \\\\Let us now specify the space-time properties of the regions in which Alice and Bob labs are located. The two space-time points $(x,x')$ belong to a space-time region $\Omega_A$ in which Alice's lab is located, while $(y,y')$ refer to points of the region $\Omega_B$ corresponding to the location of Bob's lab. The two regions $(\Omega_A,\Omega_B)$ are space-like separated. Moreover, we consider events within $\Omega_A$ which are time-like. The same for those belonging to $\Omega_B$. This means that the measurements performed by Alice and Bob are separated by space-like intervals, implementing thus the principle of relativistic causality. In summary, we have: 
\begin{equation} 
(x-x')^2 >0\;, \qquad (y-y')^2>0\;, \qquad (x-y)^2<0\;,\qquad (x-y')^2<0 \;, \qquad (x'-y)^2<0 \;, \qquad (x'-y')^2<0 \;, \label{spt}
\end{equation}
where 
\begin{equation}
(x-x')^2 = ((x^0 - x'^0)^2 ) - ({\vec x}-{\vec x'})^2 \;. \label{metric}
\end{equation}
The physical meaning of eq.\eqref{spt} can be easily visualized with the help of a two-dimensional $(t,x)$ spacetime diagram, see Fig.\eqref{contorno_c}. Alice's lab is located at $x=0$, while Bob's lab at $x=x_B$. Alice performs a first measurement at the time $t_A$ and repeats it at $t'_A > t_A$. On the other hand, Bob does his first measurement at $t_B$ and the second one at $t'_B>t_B$. Moreover, as it is apparent from  Fig.\eqref{contorno_c}, since the spatial distance between the two labs is greater than the maximum time interval, {\it i.e.} $x_B > (t'_B-t_A)$, it follows that Alice and Bob are space-like separated, according to eq.\eqref{spt}. 

\begin{figure}[!ht] 

	\includegraphics[scale=0.5]{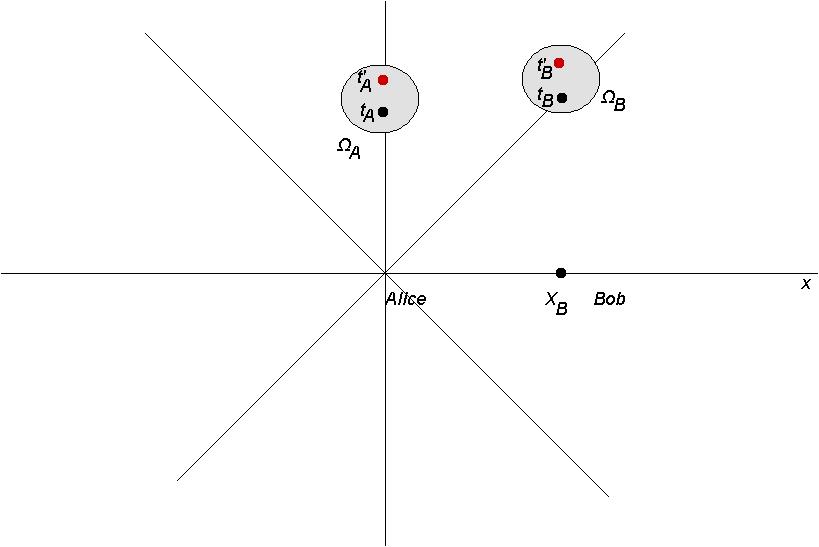}
	\caption{Location of the labs of Alice and Bob in a two-dimensional spacetime diagram }
	\label{contorno_c}
\end{figure}

\section{Canonical quantization and introduction of the $CHSH$ correlator by means of Weyl operators} \label{canon}
Before facing the  quantization of the operator \eqref{uu}, it is useful to shortly remind a few basic properties of the canonical quantization of a free massive scalar field \cite{Haag:1992hx}. For such a purpose, the use of a single field $\varphi$ is enough, the generalization to two free fields being immediate. We start  with a free Klein-Gordon field 
\begin{equation} 
{\cal L} =   \frac{1}{2} \left( \partial^\mu \varphi \partial_\mu \varphi - m^2 \varphi^2 \right) \;.  \label{cnq1}
\end{equation} 
Expanding $\varphi$ in terms of annihiliation and creation operators, we get 
\begin{equation} 
\varphi(t,{\vec x}) = \int \frac{d^3 {\vec k}}{(2 \pi)^3} \frac{1}{2 \omega(k,m)} \left( e^{-ikx} a_k + e^{ikx} a^{\dagger}_k \right) \;, \qquad k^0= \omega(k,m) = \sqrt{{\vec{k}}^2 + m^2}  \;, \label{qf}
\end{equation} 
where 
\begin{equation} 
[a_k, a^{\dagger}_q] = (2\pi)^3 2\omega(k,m) \delta^3({\vec{k} - \vec{q}}) \;, \qquad [a_k, a_q] = 0\;, \qquad [a^{\dagger}_k, a^{\dagger}_q] =0\;, \label{ccr}
\end{equation}
implementing the canonical commutation relations. A quick computation shows that
\begin{equation} 
\left[ \varphi(x) , \varphi(y) \right] = i \Delta_{\textrm{PJ}}^m (x-y) = 0 \; \qquad {\rm for } \; \;\; (x-y)^2<0 \;, \label{caus} 
\end{equation}
where $\Delta_{\textrm{PJ}}^m(x-y) $ is the Lorentz  invariant causal Pauli-Jordan function, encoding the principle of relativistic causality 
\begin{itemize} 
\item 
\begin{equation}
 \Delta_{\textrm{PJ}}^m(x-y) = \frac{1}{i} \int \frac{d^4k}{(2\pi)^3} (\theta(k^0) - \theta(-k^0)) \delta(k^2-m^2) e^{-ik(x-y)}\;, \label{it1}
\end{equation} 
\item 
\begin{equation} 
\Delta_{\textrm{PJ}}^m(x-y) = - \Delta_\textrm{{PJ}}^m(y-x) \;, \qquad (\partial^2_x + m^2) \Delta_{\textrm{PJ}}^m(x-y) = 0 \;, \label{it2}
\end{equation} 
\item
\begin{equation} 
\Delta_{\textrm{PJ}}^m(x-y) = \left( \frac{\theta(x^0-y^0) -\theta(y^0-x^0)}{2\pi} \right) \left( -\delta((x-y)^2)  +m \frac{\theta((x-y)^2)  J_1(m\sqrt{(x-y)^2}) }{ 2 \sqrt{(x-y)^2}}\right) \;,
\end{equation} 
where $J_1$ is the Bessel function. 
\end{itemize} 
However, as it stands, expression  \eqref{qf} is a too singular object, being in fact an operator valued distribution in Minkowski space \cite{Haag:1992hx}. To give a well defined meaning to eq.\eqref{qf}, one introduces the smeared field 
\begin{equation} 
\varphi(h) = \int d^4x \;\varphi(x) h(x) \;, \label{sm} 
\end{equation} 
where $h(x)$ is a test function belonging to the Schwartz space ${\cal S}(\mathbb{R}^4)$, {\it i.e.} to the space of smooth infinitely differentiable functions decreasing as well as their derivatives faster than any power of $(x) \in \mathbb{R}^4$ in any direction. The support of $h(x)$, $supp_h$, is the region in which the test function $h(x)$ is non-vanishing. Introducing the Fourier transform of  $h(x)$\footnote{It is well-known that the Fourier transform ${\hat h}(p)$ of a test function $h(x) \in {\cal S}(\mathbb{R}^4)$ is again a rapidly decreasing function, namely:  ${\hat h}(p) \in {\cal S}(\mathbb{R}^4)$. }
\begin{equation}
{\hat h}(p) = \int d^4x \; e^{ipx} h(x)  \;, \label{fft}
\end{equation} 
expression \eqref{sm} becomes 
\begin{equation} 
\varphi(h) = \int \frac{d^3 {\vec k}}{(2 \pi)^3} \frac{1}{2 \omega(k,m)} \left( {\hat h}^{*}(\omega(k,m),{\vec k}) a_k + {\hat h}(\omega(k,m),{\vec k}) a^{\dagger}_k \right)  = a_h + a^{\dagger}_h\;, \label{smft} 
\end{equation}
where $(a_h,a^{\dagger}_h)$ stand for 
\begin{equation} 
a_h = \int \frac{d^3 {\vec k}}{(2 \pi)^3} \frac{1}{2 \omega(k,m)}  {\hat h}^{*}(\omega(k,m),{\vec k}) a_k \;, \qquad 
a^{\dagger}_h = \int \frac{d^3 {\vec k}}{(2 \pi)^3} \frac{1}{2 \omega(k,m)} {\hat h}(\omega(k,m),{\vec k}) a^{\dagger}_k \;. 
\end{equation} 
One sees thus that the smearing procedure has turned the too singular object $\varphi(x)$, eq.\eqref{qf}, into an operator, eq.\eqref{smft}, acting on the Hilbert space of the system. When rewritten in terms of the operators 
$(a_h,a^{\dagger}_{h'})$, the canonical commutation relations \eqref{ccr} take the form 
\begin{equation} 
\left[ a_h,a^{\dagger}_{h'}\right]  = \langle h | h' \rangle_m \;, \label{ccrfg}
\end{equation}
where $ \langle h | h' \rangle_m$ is the Lorentz invariant scalar product between the test functions $h$ and $h'$. {\it i.e.} 
\begin{equation} 
\langle h | h' \rangle_m = \int \frac{d^3 {\vec k}}{(2 \pi)^3} \frac{1}{2 \omega(k,m)}  {\hat h}^{*}(\omega(k,m),{\vec k}) {\hat h'}(\omega(k,m), {\vec k}) = 
\int \frac{d^4 {\vec k}}{(2 \pi)^4} (2\pi \;\theta(k^0) \delta(k^2-m^2) ) {\hat h}^{*}(k) {\hat h}(k)  \;. \label{scpd}
\end{equation} 
We indicate by a subscript $m$ the mass that appears in the scalar product \eqref{scpd}, in view of the fact that our model, eq.\eqref{clact}, contains  more than one mass. The scalar product \eqref{scpd} can be rewritten in configuration space. Taking the Fourier transform, one has 
\begin{equation} 
\langle h | h' \rangle_m = \int d^4x d^4x'\; h(x) {\cal D}(x-x') h(x')  \;, \label{confsp} 
\end{equation} 
where ${\cal D} ^m(x-x')$ is the so-called Wightman function
\begin{equation} 
{\cal D}^m(x-x') = \langle 0| \varphi(x) \varphi(x') |0 \rangle = \int \frac{d^3 {\vec k}}{(2 \pi)^3} \frac{1}{2 \omega(k,m)} e^{-ik(x-x')}  \;, \qquad k^0=\omega(k,m) \;. \label{Wg}
\end{equation} 
which can be decomposed as 
\begin{equation} 
{\cal D}^m(x-x') = \frac{i}{2} \Delta_{\textrm{PJ}}^m(x-x')   + H^m(x-x') \;, \label{decomp}
\end{equation} 
where $\Delta_{\textrm{PJ}}^m(x-x') $ is the Pauli-Jordan function and $H^m(x-x')=H^m(x'-x)$ is the real symmetric quantity \cite{Scharf} 
\begin{equation} 
H^m(x-x') = \frac{1}{2} \int \frac{d^3 {\vec k}}{(2 \pi)^3} \frac{1}{2 \omega(k,m)} \left( e^{-ik(x-x')} + e^{ik(x-x')}
 \right) \qquad k^0=\omega(k,m)\;. \label{H}
\end{equation} 
The commutation relation (\ref{caus}) can be expressed in terms of smeared fields as
\begin{equation} 
\left[ \varphi(h) , \varphi(h') \right] =i  \Delta_{\textrm{PJ}}^m(h,h') \label{comm_smeared}
\end{equation}
where $h$, $h'$ are test functions and
\begin{equation} 
\Delta_{\textrm{PJ}}^m(h,h')= \int d^4x \; d^4x' h(x) \Delta_{\textrm{PJ}}^m(x-x')  h'(x'). \label{pauli_jordan_smeared}
\end{equation}
Therefore, the causality condition in terms of smeared fields becomes
\begin{equation} 
\left[ \varphi(h) , \varphi(h') \right] = 0, \label{caussm}
\end{equation}
if $supp_h$ and $supp_{h'}$ are space-like.

\subsection{A few words on the test functions}\label{testf}
As a concrete example of test functions, we might consider the class of test functions that have compact support, known as \emph{bump functions}. A good example of a bump function is the function
\begin{equation}
f_{bump}(x) = 
     \begin{cases}
       {\cal C}\;e^{-\frac{1}{\alpha^2 - m^2|x|^2} }&\quad\text{if} \; \alpha^2 \ge m^2 |x|^2 \\
       0 &\quad\text{if} \;  \alpha^2 <  m^2 |x|^2 \;,  \label{bump}
     \end{cases} 
    \end{equation}
where $\alpha$ is a real number, $\cal C$ is a normalization factor and $|x|^2 = (x^0)^2 + (x^1)^2 + (x^2)^2 + (x^3)^2$ is the Euclidean distance from the origin. The function \eqref{bump} is a smooth function, infinitely differentiable, with compact support. It is non-vanishing only within the region $ \alpha^2 \ge m^2 |x|^2 $. Bump functions as that in equation \eqref{bump} have many interesting properties, see \cite{Scharf}. Since $f_{bump}(x)= f_{bump}(-x)$, its Fourier transform 
\begin{equation} 
{\hat f}_{bump}(p) = \int d^4x \; e^{ipx} f_{bump} (x)  \;, \label{ffb}
\end{equation} 
is a real symmetric function 
\begin{equation} 
{\hat f}_{bump}(p)^{*} = {\hat f}_{bump}(p) \;, \qquad {\hat f}_{bump}(p) = {\hat f}_{bump}(-p) \;. \label{pfb}
\end{equation} 
It turns out that ${\hat f}_{bump}(p)$ has no compact support. However, it is a smooth, infinitely differentiable function, exhibiting an exponential decay for large $|p|$. As such, both $f_{bump}(x)$ and its Fourier transform, 
${\hat f}_{bump}(p)$, belong to the Schwartz space ${\cal S}(\mathbb{R}^4)$. Another important property of the bump functions is that their derivatives are still bump functions. For example 
\begin{equation}
f'_{bump}(x) = 
     \begin{cases}
       {{\cal C'}\;\frac {\partial \left( e^{-\frac{1}{\beta^2 - m^2|x|^2} }\right) }{\partial x^0} }&\quad\text{if} \; \beta^2 \ge m^2 |x|^2 \\
       0 &\quad\text{if} \;  \beta^2 <  m^2 |x|^2 \;,  \label{bump}
     \end{cases} 
    \end{equation}
is an antisymmetric bump function: $f'_{bump}(x)=- f'_{bump}(-x)$. As a consequence, its Fourier transform reads
\begin{equation} 
{\hat f'_{bump}} (p) = -ip^0 \;{\cal C'} \int_{\beta \ge m|x|} d^4x \; e^{ipx} e^{-\frac{1}{\beta^2 - m^2|x|^2}} \;. \label{antb}
\end{equation} 
We see thus that ${\hat f'_{bump}} (p)$ is a purely imaginary function which is antisymmetric: 
\begin{equation} 
{\hat f'_{bump}} (p)^{*} = -  {\hat f'_{bump}} (p)  \;,  \qquad {\hat f'_{bump}} (p) = - {\hat f'_{bump}} (-p) \label{antftb}
\end{equation} 
In particular, the normalization constants $({\cal C},{\cal C'})$ as well as the properties \eqref{pfb} and 
\eqref{antftb}, can be used to define a pair of test functions $(f,f')$  fulfilling the following properties, see also \cite{Summers:1987fn,Summ,Summers:1987ze,Summers:1988ux,Summers:1995mv}: 
\begin{eqnarray} 
f(x) &= & f(-x) \;, \qquad  f'(x)=-f'(-x) \;, \nonumber \\
{\lVert f\rVert}_m^2 & = & \langle f | f \rangle_m = m^2\;, \qquad {\lVert f'\rVert}_m^2 = m^2 \;, \nonumber \\
\langle f | f'\rangle_m & = & {\rm purely \; imaginary} = \frac{i}{2}   \int d^4x d^4x' f(x) \Delta_{\textrm{PJ}}^m(x-x') f'(x') \;. \label{pi}
\end{eqnarray}
The appearance of the mass parameter $m^2$ in the normalization of the test functions, eq.\eqref{pi}, is due to our conventions for the engineering dimensions of the quantities $(\hvarphi_A^i, \hvarphi_B^i)$, eq.\eqref{rescaled}, which will be kept dimensionless throughout. Moreover, as we shall see in the next section, the final dependence of the quantum $CHSH$ correlator from the  parameters $(m^2_A,m^2_B)$ will enable us to discuss the zero mass limit. \\\\Finally, we recall the Cauchy-Schwartz inequality for the scalar product
\begin{equation} 
\l| \langle f | f' \rangle _m\l|^2 \le {\lVert f\rVert}_m^2 \; {\lVert f'\rVert}_m^2 = m^4 \;. \label{CS}
\end{equation} 
\subsection{Weyl operators} \label{Weyl}
Let us now remind a few features of the so-called Weyl operators, which will be the building blocks for the construction of the $CHSH$ operator, eqs.\eqref{uu},\eqref{cos}, at the quantum level.  The Weyl operators are bounded unitary operators built out by exponentiating the smeared field, namely 
\begin{equation} 
{\cal A}_h = e^{i \hat{\varphi}(h) }\;, \label{Weyl}
\end{equation}
where $\hat{\varphi}(h)=\varphi (h)/m$ is the dimensionless smeared field defined in eqs.\eqref{sm}, \eqref{smft}. Making use of the following relation
\begin{equation}
e^A \; e^B = \; e^{ A+B +\frac{1}{2}[A,B] } \;, \label{exp_AB}
\end{equation} 
valid for two operators $(A,B)$ commuting with $[A,B]$, one immediately checks that the Weyl operators give rise to the following algebraic structure
\begin{equation}
{\cal A}_h \;{\cal A}_h'= e^{- \frac{1}{2} [\hat{\varphi}(h), \hat{\varphi}(h')] }\;{\cal A}_{(h+h')} = e^{ - \frac{i}{2m^2} \Delta_{\textrm{PJ}}^m(h,h')}\;{\cal A}_{(h+h')}  \;. \label{algebra} 
\end{equation} 
where $\Delta_{\textrm{PJ}}^m(h,h')$ is the causal Pauli-Jordan function, eq.\eqref{caussm}. Also, using the canonical commutation relations written in the form \eqref{ccrfg}, for the vacuum expectation value of ${\cal A}_h$, one gets 
\begin{equation} 
\langle 0| \; {\cal A}_h \; |0 \rangle = \; e^{-\frac{1}{2m^2} {\lVert h\rVert}_m^2 } \;, \label{vA}
\end{equation} 
As already underlined, the vacuum state $|0>$ is the Fock vacuum: $a_k|0>=0$, for all modes $k$. 

\subsection{Construction of the CHSH quantum operator}\label{CHSH_quantum} 
At the classical level, the model we are  considering is characterized  by the Lagrangian density (\ref{clact}). We have two free fields $(\varphi_A^i,\varphi_B^i)$, for Alice and Bob, respectively.  Each field satisfies the Klein-Gordon equation and can be expanded in term of annihilation and creation operators, eq.(\ref{qf}), namely
\begin{eqnarray} 
\varphi_A^i(t,{\vec x}) & = & \int \frac{d^3 {\vec k}}{(2 \pi)^3} \frac{1}{2 \omega(k,m_A)} \left( e^{-ikx} a^i_k + e^{ikx} a^{i\,\dagger}_k \right) \;, \qquad k^0= \omega(k,m_A)  \;, \nonumber \\
\varphi_B^i(t,{\vec x}) & = & \int \frac{d^3 {\vec k}}{(2 \pi)^3} \frac{1}{2 \omega(k,m_B)} \left( e^{-ikx} b^i_k + e^{ikx} b^{i\,\dagger}_k \right) \;, \qquad k^0= \omega(k,m_B)  \;,\label{alice_bob_fields}
\end{eqnarray} 
where the only non-vanishing commutators among the annihilation and creation operators are
\begin{eqnarray} 
\left[ a_k^i, a^{j\dagger}_q\right]  & = &(2\pi)^3 2\omega(k,m_A) \delta^3({\vec{k} - \vec{q}})\delta^{ij}\, , \nonumber \\
\left[ b_k^i, b^{j\dagger}_q\right]  & = &(2\pi)^3 2\omega(k,m_B) \delta^3({\vec{k} - \vec{q}})\delta^{ij}\, .
 \label{ccr_alice_bob}
\end{eqnarray}
To have well defined operators in the Fock-Hilbert space, these fields are smeared with test functions, as described in (\ref{sm}), resulting in $\left(\varphi_A^i(h),\;\varphi_B^i(h)\right)$. It is thus  straightforward to evaluate the following commutation relations for the smeared fields
\begin{eqnarray} 
	\left[ \varphi^i_A(h) , \varphi^j_A(h') \right] &=& i\delta^{ij} \Delta_{\textrm{PJ}}^{m_A}(h,h')\; , \nonumber \\
	\left[ \varphi^i_B({\tilde h}) , \varphi^j_B({\tilde h'}) \right] &=& i\delta^{ij} \Delta_{\textrm{PJ}}^{m_B}({\tilde h},{\tilde h}')\;, \nonumber \\
	\left[ \varphi^i_A(h) , \varphi^j_B({\tilde h}) \right] &=& 0\;,
		 \label{caussm}
\end{eqnarray}
valid for any pair of test functions $(h,h')$, $({\tilde h}, {\tilde h'})$. The presence of the Pauli-Jordan function in expressions \eqref{caussm}  implements the relativistic causality  in the model. In fact, if $supp_h$ and $supp_{h'}$ are space-like as well as those of $({\tilde h},{\tilde h'})$, then the commutator of the corresponding  smeared fields vanishes.  The Fock vacuum  of the model  is defined as being the state $| 0 \rangle$ such that
\begin{eqnarray}
a^i_{k}|0 \rangle=0\, , \nonumber \\
b^i_{k}|0 \rangle=0\, , \label{fock_vacuum_a_b}
\end{eqnarray}
for any $i=1,2,3$ and any momentum $k$. \\\\We are now ready to write down the quantum version of the $CHSH$ operator, eqs.\eqref{uu},\eqref{cos}. We first introduce the smeared operator 
\begin{eqnarray} 
{\cal{U}}^{ab}_{ff'gg'}  & =&  \cos a^i(\hvarphi_A^i(f) + \hvarphi_B^i(g) ) \; \cos b^i(\hvarphi_A^i(f') + \hvarphi_B^i(g') )  \; \nonumber \\
&=& \left[  \frac{ e^{ia^i(\hvarphi_A^i(f) + \hvarphi_B^i(g) )} + e^{-ia^i(\hvarphi_A^i(f) + \hvarphi_B^i(g) )}}{2} \right]
\left[\frac{ e^{ib^i(\hvarphi_A^i(f') + \hvarphi_B^i(g') )} + e^{-ib^i(\hvarphi_A^i(f') + \hvarphi_B^i(g') }}{2} \right]\;, \label{Uab}
\end{eqnarray} 
where $(f,f')$ and $(g,g')$ are test functions belonging, respectively, to Alice and Bob space-time regions, $\Omega_A$, $\Omega_B$, see Fig.\eqref{contorno_c}. More precisely, the supports of $(f,f')$ are space-like with respect to those of $(g,g')$. 
\begin{equation} 
( supp_{(f,f')})   \;\;\;\;  {\rm space-like\; with \; respect \;to} \;\;\;\; (supp_{(g,g')}) \;. \label{supp} 
\end{equation}
Further, we introduce the Hermitian operator 
\begin{equation} 
{\cal{\hat U}}^{ab}_{ff'gg'}  =  ({\cal{\hat U}}^{ab}_{ff'gg'})^{\dagger} = \frac{1}{2} \left(  {\cal{U}}^{ab}_{ff'gg'}+ ({\cal{U}}^{ab}_{ff'gg'})^{\dagger} \right) \;. \label{HU}
\end{equation}
Finally, for quantum version of the $CHSH$ operator, eqs.\eqref{uu},\eqref{cos}, we write 
\begin{equation} 
{\cal C}^{aa'bb'}_{(ff'gg')} = {\cal{\hat U}}^{ab}_{ff'gg'}+ {\cal{\hat U}}^{a'b}_{ff'gg'} + {\cal{\hat U}}^{ab'}_{ff'gg'} -{\cal{\hat U}}^{a'b'}_{ff'gg'} \;. \label{UUf}
\end{equation} 
In the following, we shall compute the vacuum correlator 
\begin{equation}
\langle 0|\; {\cal C}^{aa'bb'}_{(ff'gg')} \; |0 \rangle \;, \label{corr}
\end{equation} 
by means of the algebraic properties of the Weyl operators. According to \cite{Summers:1987fn,Summ,Summers:1987ze,Summers:1988ux,Summers:1995mv}, we shall speak of a violation of the $CHSH$ classical inequality, eq.\eqref{le2}, if 
\begin{equation}
\l| \langle 0|\; {\cal C}^{aa'bb'}_{(ff'gg')} \; |0 \rangle \l| > 2 \;, \label{corrv}
\end{equation} 
for some suitable choice of $(a^i,a^{'i},b^{i},b^{'i})$. \\\\Expression \eqref{corr} can be calculated in closed form  using \eqref{exp_AB}-\eqref{vA} and the fact that the vacuum is annihilated by $a_h^i$ and $b_h^j$, see \eqref{fock_vacuum_a_b}. The outcome of our result reads
\begin{eqnarray}
\langle \mathcal{C}_{\left(ff'gg'\right)}^{aa'bb'} \rangle & = & e^{-\frac{1}{2}\left[\overrightarrow{a}\cdot\overrightarrow{a}\left(\frac{\left\Vert f\right\Vert _{m_{A}}^{2}}{m_A^2}+\frac{\left\Vert g\right\Vert _{m_{B}}^{2}}{m_B^2}\right)+\overrightarrow{b}\cdot\overrightarrow{b}\left( \frac{\left\Vert f'\right\Vert _{m_{A}}^{2}}{m_A^2}+\frac{\left\Vert g'\right\Vert _{m_{B}}^{2}}{m_B^2}\right)\right]}\cos\left(\frac{\overrightarrow{a}\cdot\overrightarrow{b}}{2}\left(\omega_{A}+\omega_{B}\right)\right)\cosh\left(\overrightarrow{a}\cdot\overrightarrow{b}\left(\widetilde{\omega}_{A}+\widetilde{\omega}_{B}\right)\right)\nonumber \\
&  & +\left(a\rightarrow a'\right)\nonumber \\
&  & +\left(b\rightarrow b'\right)\nonumber \\
&  & -\left(a\rightarrow a',\;b\rightarrow b'\right) \, ,\label{chsh_exact}
\end{eqnarray}
where
\begin{eqnarray}
\omega_A &=& \frac{1}{m_A^2}\Delta_{\textrm{PJ}}^{m_A}(f,f')\; , \nonumber \\
\omega_B &=& \frac{1}{m_B^2}\Delta_{\textrm{PJ}}^{m_B}(g,g')\; , \nonumber \\
\widetilde{\omega}_{A} & = & \frac{1}{m_A^2} \textrm{Re}\left\langle f|f'\right\rangle _{m_{A}},\nonumber \\
\widetilde{\omega}_{B} & = & \frac{1}{m_B^2}\textrm{Re}\left\langle g|g'\right\rangle _{m_{B}}. \label{omegas}
\end{eqnarray}
Instead of explicitly writing all terms in eq.\eqref{chsh_exact}, we have simply indicated that the other terms are obtained from the first one  by replacing the vectors $\vec{a}=(a^1,\,a^2,\,a^3\,)$ and $\vec{b}=(b^1,\,b^2, \,b^3\,)$ as denoted by the arrows. The scalar product between the vectors in \eqref{chsh_exact} is the tridimensional Euclidian scalar product, \emph{i.e},
$\vec{a}\cdot\vec{b}=\sum_{i=1}^{3}a^ib^i$.

\section{Analysis of the violation of the  CHSH inequality} \label{CHSH}
Having evaluated the $CHSH$ correlator, eq.\eqref{chsh_exact}, we can face now the issue of the violation of the $CHSH$ inequality. To a first look, one might have the impression that eq.\eqref{chsh_exact} contains a lot of free parameters, so that it would be relativily simple to find  a violation of the $CHSH$ inequality. Though, things are not that easy, the main reason being the presence of the exponentials which decay  very fast. As a consequence,  the allowed space of parameters turns out to be quite small. \\\\Before analysing the best choice for the parameters $(a^i,a^{'i},b^{i},b^{'i})$, we  fix the norms of the test functions $\left( f,f'\right)$ and $\left( g,g'\right)$, with supports in the regions of Alice's lab $\Omega_A$ and Bob's lab $\Omega_B$, respectively, according to 
\begin{equation} 
{\lVert f\rVert}_{m_A}^2   = {\lVert f'\rVert}_{m_A}^2   = m_A^2\;, \qquad {\lVert g\rVert}_{m_B}^2   = {\lVert g'\rVert}_{m_B}^2   = m_B^2\;. \label{norms}
\end{equation} 
It is worth remarking here that, as expected,  the choice of the norm of the test functions does not play much role in expression \eqref{chsh_exact} . These norms  are easily seen to be reabsorvable  into the vectors $\left( \vec{a},\,\vec{b},\,\vec{a'},\,\vec{b'}\right)$, which are arbitrary. Therefore, the choice of working with normalized test functions, eq.\eqref{norms}, does not change the final output. Concerning now the scalar products $\langle f | f'\rangle$ and $\langle g | g'\rangle$, we have followed the same prescription adopted in the original work \cite{Summ}  and have taken $(\langle f | f'\rangle,\langle g | g'\rangle)$ purely imaginary\footnote{We notice that, since at least one of the labs is not located at the origin of the coordinate system, one pair of test functions will not have the odd and even symmetries with respect to $x=0$, as assumed in \eqref{pi}. However,  due to the translation invariance of the Wightman function and, consequently, of the scalar product, a pair of test functions can still satisfy eqs.\eqref{pi} if they are odd and even with respect to a certain point of the space-time which, in the present case, is the location of Bob's lab, see Fig.\eqref{contorno_c}.}, as described in eqs.\eqref{pi}, namely 
\begin{eqnarray} 
\langle f | f'\rangle & = & {\rm purely \; imaginary} = \frac{i}{2}   \Delta_{\textrm{PJ}}^{m_A}(f,f') \;, \nonumber \\
\langle g | g'\rangle & = & {\rm purely \; imaginary} = \frac{i}{2}   \Delta_{\textrm{PJ}}^{m_B}(g,g') \;,
 \label{piii}
\end{eqnarray}
Due to eqs.\eqref{norms}, \eqref{piii},  the $CHSH$ correlator gets simplified:  
\begin{eqnarray}
\langle \mathcal{C}_{\left(ff'gg'\right)}^{aa'bb'} \rangle & = & e^{-\vec{a}\cdot\vec{a}-\vec{b}\cdot\vec{b}}\cos\left(\frac{\vec{a}\cdot\vec{b}}{2}\left(\omega_{A}+\omega_{B}\right)\right)+e^{-\vec{a'}\cdot\vec{a'}-\vec{b}\cdot\vec{b}}\cos\left(\frac{\vec{a'}\cdot\vec{b}}{2}\left(\omega_{A}+\omega_{B}\right)\right)\nonumber \\
&  & +e^{-\vec{a}\cdot\vec{a}-\vec{b'}\cdot\vec{b'}}\cos\left(\frac{\vec{a}\cdot\vec{b'}}{2}\left(\omega_{A}+\omega_{B}\right)\right)-e^{-\vec{a'}\cdot\vec{a'}-\vec{b'}\cdot\vec{b'}}\cos\left(\frac{\vec{a'}\cdot\vec{b'}}{2}\left(\omega_{A}+\omega_{B}\right)\right).
 \label{chsh_simplified_first}
\end{eqnarray}
In addition to the vectors $(a^i,a'^i,b^i,b'^{i})$, expression \eqref{chsh_simplified_first} contains the quantity
\begin{eqnarray}
\omega_A+\omega_B&=&\frac{1}{m_A^2}\Delta_{\textrm{PJ}}^{m_A}(f,f')+\frac{1}{m_B^2}\Delta_{\textrm{PJ}}^{m_B}(g,g'), \label{omega_def}
\end{eqnarray}
which is the smearing of the Pauli-Jordan function. Due to choice of the scalar products $\langle f | f'\rangle$ and $\langle g | g'\rangle$ done in  eq.\eqref{piii}, from the Cauchy-Schwarz inequality it follows 
\begin{eqnarray}
\Delta_{\textrm{PJ}}^{m_A}(f,f')&=&\frac{2}{i}\langle f | f' \rangle_{m_A}\, , \label{ff'} 
\end{eqnarray}
which implies that
\begin{equation}
	\left|\Delta_{\textrm{PJ}}^{m_{A}}\left(f,f'\right)\right| \leq  2\left\Vert f\right\Vert _{m_{A}}\left\Vert f'\right\Vert _{m_{A}}=2m_A^2. \label{ineq_ff'}
\end{equation}
The same holds for $\Delta_{\textrm{PJ}}^{m_B}(g,g')$. Thus, taking into account the even character of the cosine, it is convenient to parametrize $(\omega_A+\omega_B)$ by introducing the quantity $\sigma$ defined as:
\begin{equation}
\omega_A+\omega_B=4\sigma, \qquad 0 \leq \sigma \leq 1. \label{omega_sigma}
\end{equation}
Therefore, we have 
\begin{eqnarray}
\langle \mathcal{C}_{\left(ff'gg'\right)}^{aa'bb'} \rangle & = & e^{-a^2-b^2}\cos\left(2\,\vec{a}\cdot\vec{b}\,\sigma\right)+e^{-a'^2-b^2}\cos\left(2\,\vec{a'}\cdot \vec{b}\,\sigma\right)\nonumber \\
&  & +e^{-a^2-b'^2}\cos\left(2\,\vec{a}\cdot \vec{b'}\,\sigma\right)-e^{-a'^2-b'^2}\cos\left(2\,\vec{a'}\cdot\vec{b'}\,\sigma\right),
\label{chsh_simplified}
\end{eqnarray} 
where $a=\left| \vec{a}\right|$, $b=\left| \vec{b}\right|$, $a'=\left| \vec{a'}\right|$, $b'=\left| \vec{b'}\right|$.  Before focussing on expression \eqref{chsh_simplified}, let us devote a little discussion to the parameter $\sigma$, which has a deep physical meaning. This is the task of the next subsection.

\subsection{The meaning of the parameter $\sigma$}\label{sigsig}  
As we have seen, eq.\eqref{omega_sigma}, the parameter $\sigma$ is directly related to the smearing of the Pauli-Jordan functions $(\Delta_{\textrm{PJ}}^{m_{A}}\left(f,f'\right), \Delta_{\textrm{PJ}}^{m_{B}}\left(g,g'\right)$. As such, $\sigma$ encodes all information about the relativistic causality of our model. It is important thus to have a more precise idea of its behavior and of its explicit relation with  Alice and Bob space-time configurations, as depicted in Fig.\eqref{contorno_c}. To that end, it suffices to pick up Alice's factor $\omega_A$, eq.\eqref{omegas}, and proceed by smearing it with two narrowed Gaussians in order to  be able to work out  analytic expressions which will provide a more transparent  understanding of $\sigma$. Accordingly, for the pair of Gaussian test functions $(f,f')$ around $\overrightarrow{x}=0$, we write
	\begin{eqnarray}
	f\left(x\right) & = & \sqrt{\frac{8\pi^{2}}{0,41411\ldots}}m_A\partial_{t}e^{-m^{2}_A\left(t-t_{A}\right)^{2}}e^{-m^{2}_Ar^{2}}\,,\nonumber \\
	f'\left(x\right) & = & \sqrt{\frac{8\pi^{2}}{0,18301\ldots}}m^{2}_Ae^{-m^{2}_A\left(t-t'_{A}\right)^{2}}e^{-m^{2}_Ar^{2}},\label{eq:gaussian_test_functions-1}
	\end{eqnarray}
	where $r=\left|\vec{x}\right|$ and where the numerical factors take into account the normalization of $(f,f')$.  The two temporal coordinates $t_A$ and $t'_A$ in eq.\eqref{eq:gaussian_test_functions-1} can be thought of as the instants in which Alice performs her measurements, see Fig.\eqref{contorno_c}.	 Moving to momentum space, one gets\footnote{Strictly speaking, being $f$ and $f'$ two Gauusians, they do not display 
properties \eqref{pi}, that is $\left\langle f|f'\right\rangle$ has
a real part as well as an imaginary part. Moreover,  one easily verifies that the real part gets smaller and smaller as $m_A\left|t_{A}-t'_{A}\right|$ becomes small, so that  $\left\langle f|f'\right\rangle$ fulfills, in practice, property \eqref{pi}.}
\begin{eqnarray}
	\hat{f}\left(p\right) & = & -\sqrt{\frac{8\pi^{2}}{0,41411\ldots}}\frac{ip^{0}e^{-\frac{p^{2}}{4m^{2}}-\frac{p_{0}^{2}}{4m^{2}}+ip^{0}t_{A}}}{m\sqrt{2}}\,,\nonumber \\
	\hat{f}'\left(p\right) & = & \sqrt{\frac{8\pi^{2}}{0,18301\ldots}}\frac{e^{-\frac{p^{2}}{4m^{2}}-\frac{p_{0}^{2}}{4m^{2}}+ip^{0}t'_{A}}}{\sqrt{2}}\,.\label{eq:gaussian_momentum_space-1}
	\end{eqnarray}
Evaluating $\omega_A$, one finds
	\begin{eqnarray}
	\omega_A & = & \frac{2}{\sqrt{0,41411\ldots}\sqrt{0,18301\ldots}}\int_{0}^{\infty}du\,u^{2}\cos\left(m_A\left(t_{A}-t'_{A}\right)\sqrt{1+u^{2}}\right)e^{-\frac{1+2u^{2}}{2}}.\label{eq:omega_example}
	\end{eqnarray}
The behavior of $\omega_A$ as a function of $m_A\Delta t\equiv m_A\left|t_{A}-t'_{A}\right|$
is shown in Fig.\eqref{omega}. Essentially, $\omega_A$ shows an exponential decay modulated by a periodic function.  This exponential decay in the variable $(m_A\Delta t )$ is in full agreement with one of the main results of  \cite{Summers:1987fn,Summ,Summers:1987ze,Summers:1988ux,Summers:1995mv}, see in particular Corollary 4.2 of \cite{Summers:1987fn}. It means that the violation of the $CHSH$ inequality decreases exponentially with the magnitude of the masses of the particles and with the size of the time intervals involved. For lighter particles and short time intervals, {\it i.e. } when $\sigma \approx 1$, we shall in fact be able to show that the violation of the $CHSH$ inequality of our model is the biggest one. 
	\begin{figure}[!ht] 

		\centering
		\includegraphics[scale=0.5]{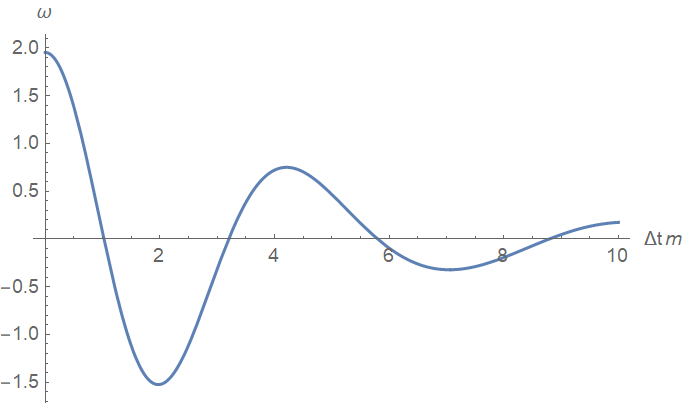}
		\caption{$\omega_A$ as a function of the variable $(\Delta t\, m_A)$.}
		\label{omega}
	\end{figure}

\subsection{The violation of the $CHSH$ inequality}\label{violationCHSH}

Let us now dive into the analysis  of the $CHSH$ correlator \eqref{chsh_simplified}, which is recognized to be a  bounded quantity. To analyze $\langle {\cal{C}}_{\left(ff'gg'\right)}^{aa'bb'} \rangle $, we consider the space of parameter as being 
\begin{equation*}
(a,\,a',\,b, \, b', \,\alpha,\,\beta, \,\gamma ,\, \delta,\,\sigma)\,,
\end{equation*}
where 
\begin{eqnarray}
\vec{a}\cdot\vec{b}&=&a \,b\cos \alpha \, , \nonumber \\
\vec{a'}\cdot\vec{b}&=& a'\,b\cos \beta \, , \nonumber  \\
\vec{a}\cdot\vec{b'}&=&a\,b'\cos \gamma \, , \nonumber \\
\vec{a'}\cdot\vec{b'}&=&a'\,b'\cos \delta.
\end{eqnarray}
Regarding the parameters $(a,\,a',\,b,\,b')$, which corresponds to the norms of the vectors $(a^i,\,a'^{i},\,b^{i},\,b'^{i})$, one sees that,  due to the exponential decay of expression \eqref{chsh_simplified}, they cannot take large values, otherwise the whole correlator will be exponentially suppressed, becoming  too small in order to detect a violation of the $CHSH$ inequality. For the two parameters $(a,b)$ the best values seem to be: $a=b=0,001$. In Fig.\eqref{violation_figure1} and Fig.\eqref{violation_figure2} one finds the behavior of $\langle {\cal{C}}_{\left(ff'gg'\right)}^{aa'bb'}\rangle $ as a function of the remaining parameters $a'$ and $b'$, and for the choices  $\sigma=0,85$ and $\sigma=1$, respectively, in a configuration in which  all vectors $(a^i,\,a'^{i},\,b^{i},\,b'^{i})$ are parallel, that is $\alpha=\beta=\gamma=\delta=0$. \\\\It turns out that, see cyan surface and the blue curve of Fig.\eqref{violation_figure1} and Fig.\eqref{violation_figure2},  $\langle {\cal{C}}_{\left(ff'gg'\right)}^{aa'bb'}\rangle $  violates the $CHSH$ inequality in both cases, although the violation is rather small, its maximum value being located in the interval $[2.029, 2.03]$.  According to \cite{Summers:1987fn} and as discussed before,  the violation of the $CHSH$  inequality reaches its  optimal value for $\sigma=1$, corresponding to light particles and short time intervals.\\\\Let us   mention that, although  in our analysis we have employed many different configurations for the vectors $(\vec{a},\,\vec{b},\,\vec{a'},\,\vec{b'})$,  the correlator $\langle {\cal{C}}_{\left(ff'gg'\right)}^{aa'bb'} \rangle$ turns out to be  sensible only  to the relative orientation of $\vec{a'}$ and $\vec{b'}$. For instance, if $\vec{a'}$ and $\vec{b'}$ are perpendicular, {\it i.e.}  $\delta=\pi/2$, the violation no longer occurs, whereas parallel,  $\delta=0$, or anti-parallel, $\delta=\pi$, configurations lead to a violation. \\\\Finally, as already underlined, the violation of the $CHSH $ inequality increases as $\sigma \rightarrow 1$. 
\begin{figure}[h!]
	\centering
	\begin{subfigure}[b]{0.4\linewidth}
		\includegraphics[width=\linewidth]{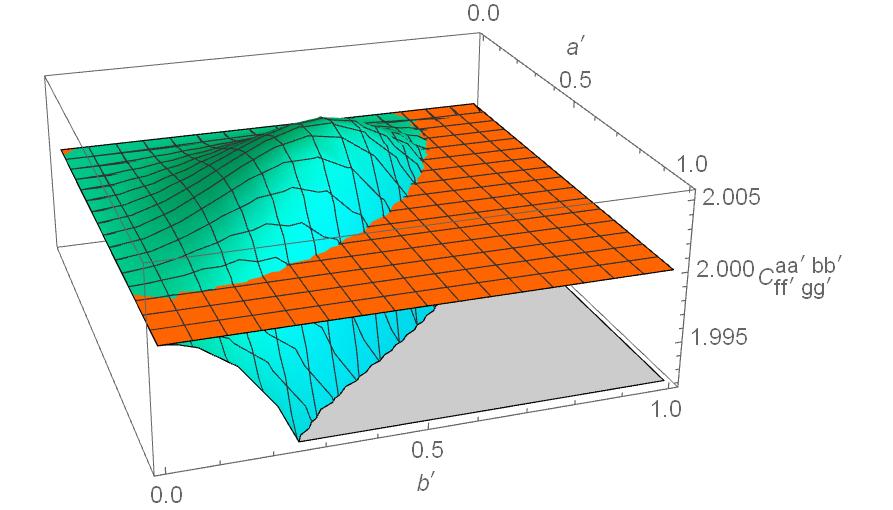}
		\end{subfigure}
		\begin{subfigure}[b]{0.4\linewidth}
		\includegraphics[width=\linewidth]{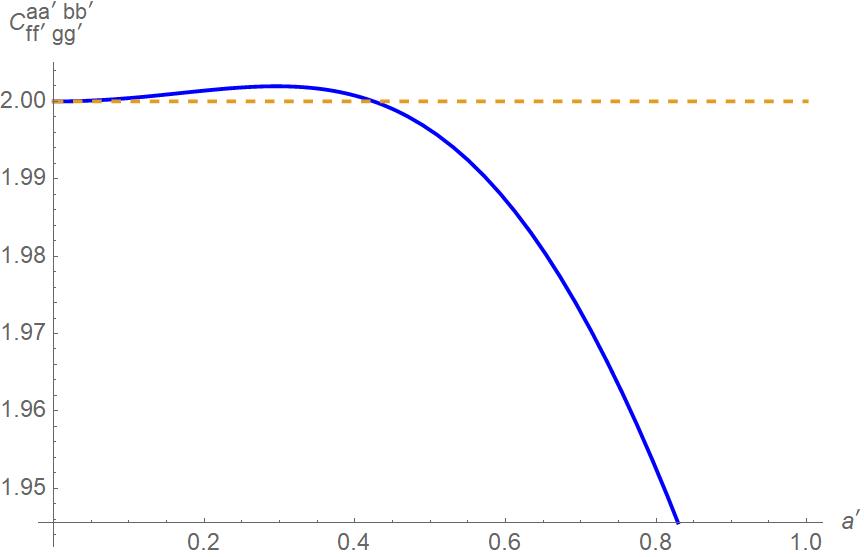}
		\end{subfigure}\\[0,5cm]
\caption{ {\bf $CHSH$  correlator for $\sigma = 0,85$}.\\[2mm] {\it Behavior of the $CHSH$ correlator $\langle {\cal C}^{aa'bb'}_{(ff'gg')} \rangle $, cyan surface, for $\vec{a}\cdot\vec{b}=\vec{a}\cdot\vec{b'}=\vec{a'}\cdot\vec{b}=0$, $\vec{a'}\cdot\vec{b'}=a'b'$, $a=b=0,001$ 
and $\sigma=0,85$. To observe the violation more easily, we have also plotted the plane z=2, corresponding to the  orange surface. The blue line in the right hand side figure shows the behavior of  $\langle {\cal C}^{aa'bb'}_{(ff'gg')}\rangle $ for $\vec{a}\cdot\vec{b}=\vec{a}\cdot\vec{b'}=\vec{a'}\cdot\vec{b}=0$, $\vec{a'}\cdot\vec{b'}=a'b'$, $a=b=0,001$, $\sigma=0,85$ and $b'=0,7$.}}
	\label{violation_figure1}
\end{figure} 
\begin{figure}[h!]
	\centering
	\begin{subfigure}[b]{0.4\linewidth}
		\includegraphics[width=\linewidth]{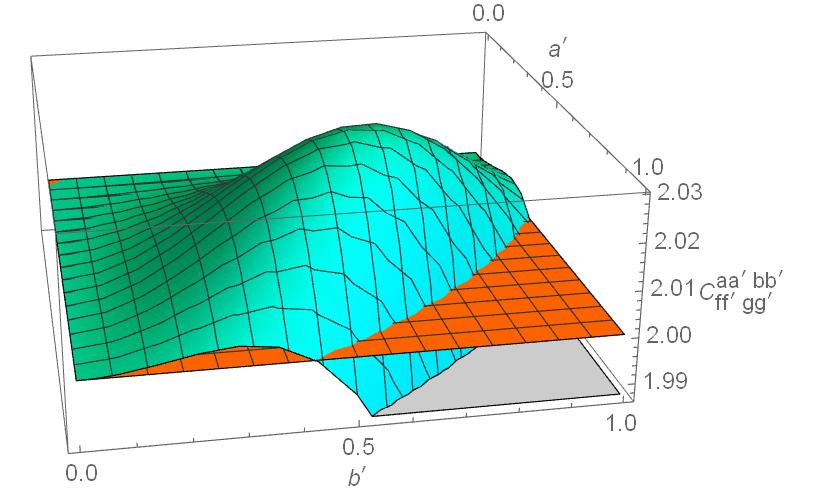}
		\end{subfigure}
\begin{subfigure}[b]{0.4\linewidth}
		\includegraphics[width=\linewidth]{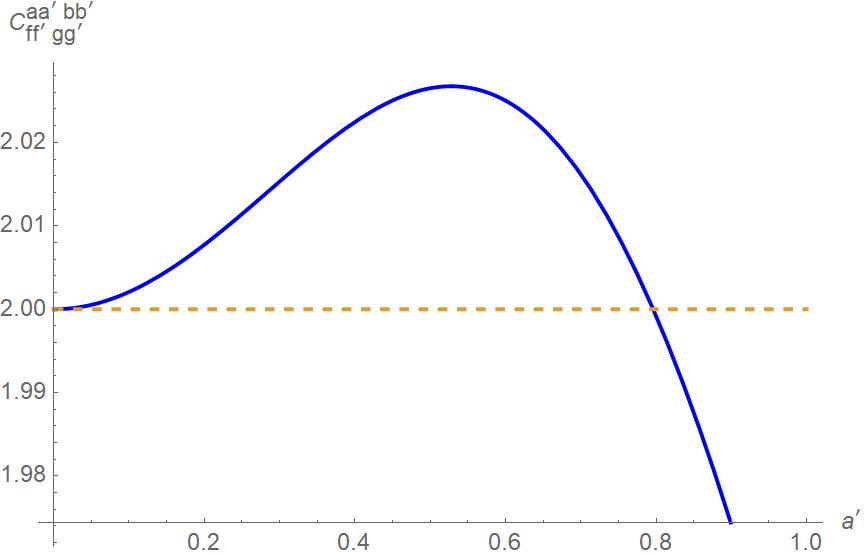}
		\end{subfigure}\\[0,5cm]
\caption{ {\bf CHSH correlator for $\sigma = 1$.} \\[2mm]
{\it The $CHSH$ correlator $\langle {\cal C}^{aa'bb'}_{(ff'gg')}\rangle$,  cyan surface,  for $\vec{a}\cdot\vec{b}=\vec{a}\cdot\vec{b'}=\vec{a'}\cdot\vec{b}=0$, $\vec{a'}\cdot\vec{b'}=a'b'$, $a=b=0,001$ and $\sigma=1$. To observe the violations more easily, we have also  plotted the plane z=2,  orange surface. The blue line in the right hand side figure shows the behavior of  $\langle {\cal C}^{aa'bb'}_{(ff'gg')}\rangle $ for $\vec{a}\cdot\vec{b}=\vec{a}\cdot\vec{b'}=\vec{a'}\cdot\vec{b}=0$, $\vec{a'}\cdot\vec{b'}=a'b'$, $a=b=0,001$, $\sigma=1$ and $b'=0,7$.}}
	
		\label{violation_figure2}
\end{figure} 

\vspace{5cm}

\section{A $BRST$ invariant formulation of the $CHSH$ inequality in gauge theories: the example of the $U(1)$ Higgs model} \label{Hg}
This section is devoted to outline a $BRST$ invariant setup for the study of the $CHSH$ inequality in gauge theories, taking as explicit example the renormalizable $U(1)$ Higgs model whose action, including the gauge fixing, is given by 
\begin{eqnarray}
S_{\textrm{Higgs}} & = & \int d^{4}x\left[-\frac{1}{4}F_{\mu\nu}(A)F^{\mu\nu}(A)+\left(D_{\mu}\varphi\right)^{\ast}\left(D^{\mu}\varphi\right)-\frac{\lambda}{2}\left(\varphi^{\ast}\varphi-\frac{v^{2}}{2}\right)^{2}+b\partial_{\mu}A^{\mu}+\overline{c}\partial^{2}c\right] \;, \label{higgs_action}
\end{eqnarray}
where 
\begin{eqnarray}
\varphi & = & \frac{1}{\sqrt{2}}\left(v+h+i\rho\right) \label{phi_para}
\end{eqnarray}
is a complex scalar field whose components $h$ and $\rho$ denote the Higgs and the Goldstone fields, respectively. The massive parameter $v$ is the $vev$ of $\varphi$, {\it i.e} $\langle \varphi \rangle= \frac{v}{\sqrt{2}}$, implementing the Higgs mechanism. The field $b$ is known as the Nakanishu-Lautrup field, needed to impose the gauge condition which, in the present case, has been chosen to be the transverse Landau gauge
\begin{equation} 
\partial A=0 \;. \label{lgauge}
\end{equation}
Also, the fields $(\overline{c},c)$ are the Faddeev-Popov ghosts. \\\\The action $S_{\textrm{Higgs}}$ enjoys an exact $BRST$ invariance:
\begin{equation} 
sS_{\textrm{Higgs}} = 0 \;, \qquad s^2=0 \;, \label{exacts}
\end{equation}
where $s$ is the nilpotent $BRST$ operator, whose action on the fields $(A_\mu,h,\rho,b, \overline{c},c)$ is specified by 
\begin{eqnarray}
sA_{\mu} & = & -\partial_{\mu}c\,,\nonumber \\
sh & = & -ec\rho\,,\nonumber \\
s\rho & = & ec\left(v+h\right)\,,\nonumber \\
sc & = & 0\,,\nonumber \\
s\overline{c} & = & b\,,\nonumber \\
sb & = & 0\,.  \label{BRST_transf}
\end{eqnarray}
The interest in the $U(1)$ Higgs model is due to a set of  articles \cite{hooft,Frohlich:1980gj,Frohlich:1981yi,Dudal:2021dec,Dudal:2021pvw,Dudal:2020uwb,Capri:2020ppe,Dudal:2019pyg,Dudal:2019aew,Maas:2019nso,Maas:2017xzh,Sondenheimer:2019idq} where a fully $BRST$ invariant description of the massive gauge boson has been worked out. \\\\More precisely, it has been shown that the following dimension three vector operator 
\begin{equation}
V_{\mu}=\frac{1}{2}\left(-\rho \partial_{\mu} h + (v+h)\partial_{\mu} \rho + e A_{\mu} (v^2+h^2+2v h+\rho^2) \right) \;, \label{V_op}
\end{equation}
displays the following properties, see \cite{Dudal:2021dec,Dudal:2021pvw,Dudal:2020uwb,Capri:2020ppe,Dudal:2019pyg,Dudal:2019aew}:
\begin{itemize} 
\item $V_\mu$ is $BRST$ invariant, belonging to the local cohomolgy \cite{Piguet:1995er} of the operator $s$
\begin{equation} 
s V_\mu =0\;, \qquad V_\mu \neq sQ_\mu \;, \label{coh}
\end{equation} 
for some local field polynomial $Q_\mu$. 
\item $V_\mu$ turns out to be the conserved Noether current corresponding to the global $U(1)$ invariance of the action \eqref{higgs_action}, namely 
\begin{equation} 
\delta h =- \omega \rho\;, \qquad \delta \rho = \omega (v+h)\;, \qquad \delta(A_\mu,b, \overline{c},c)=0\;, \qquad \delta S_{\textrm{Higgs}} =0 \;, \label{glu1}
\end{equation}
where $\omega$ is a constant parameter. Thus
\begin{equation} 
\partial^{\mu} V_\mu = {\rm eqs.\; of\; motion} \label{Noether}
\end{equation}
From this property, it follows that the anomalous dimension of $V_\mu$ vanishes to all orders in perturbation theory. 
\item The  transverse component of the two-point function $\langle V_\mu(p) V_\nu(-p) \rangle^T $
\begin{equation} 
\langle V_\mu(p) V_\nu(-p) \rangle^T = {\cal P}_{\mu\sigma} \langle V^\sigma(p) V_\nu(-p) \rangle \;, \qquad 
{\cal P}_{\mu\sigma} =\left( g_{\mu\sigma} - \frac{ p_\mu p_\sigma}{p^2} \right)  \;, \label{trs}
\end{equation} 
has the same pole mass of the elementary two-point function $\langle A_\mu(p) A_\nu(-p) \rangle$, a key property which extends to all orders of perturbation theory, due to a set of Ward identities. 

\item The longitudinal component of $\langle V_\mu(p) V_\nu(-p) \rangle^L $
\begin{equation} 
\langle V_\mu(p) V_\nu(-p) \rangle^L = {\cal L}_{\mu\sigma} \langle V^\sigma(p) V_\nu(-p) \rangle \;, \qquad 
{\cal L}_{\mu\sigma} =\left( \frac{ p_\mu p_\sigma}{p^2} \right)  \;, \label{lrs}
\end{equation}
has only tree level contributions to all orders. Moreover, the tree level term is momentum independent, so that $\langle V_\mu(p) V_\nu(-p) \rangle^L $ does not correspond to any propagating mode. 
\item the two-point transverse function $\langle V_\mu(p) V_\nu(-p) \rangle^T$ exhibits a K{\"a}ll{\'e}n-Lehmann spectral representation with positive definite spectral density. 
\end{itemize} 
All these non-trivial properties enable us to employ the  operator $V_\mu$ to achieve a fully gauge invariant description of the massive gauge boson in the $U(1)$ Higgs model. It worth underlining that the whole set of properties listed above generalize to the non-Abelian $SU(2)$ case with a single scalar field in the fundamental representation, see \cite{Dudal:2021dec,Dudal:2021pvw,Dudal:2020uwb,Capri:2020ppe,Dudal:2019pyg,Dudal:2019aew}. \\\\Being $BRST$ invariant, the operator $V_\mu$ leads to a natural construction of $BRST$ invariant bounded Weyl type operators, {\it i.e.} 
\begin{equation} 
{\cal A}_V  =  e^{i {\hat V(f)}}  \label{Weyl_V} = e^{i \int_\Omega d^4x \; f_\mu(x) {\hat V(x)^{\mu} }}
\end{equation} 
where  ${\hat V}_\mu$ stands for the  dimensionless quantity 
\begin{equation} 
{\hat V}_\mu(x) = \frac{1}{e v^3} V_\mu (x)\;, \label{dv}
\end{equation} 
and where $\{f_\mu(x) \}$ are a set of smooth functions with compact suuport, introduced in order to localize the operator ${\cal A}_V$ in the desired region of the spacetime $\Omega$. \\\\It is helpful to remind here   that, in the case of a gauge field $A_\mu(x)$, the smearing procedure is done by means of a set $\{ f_\mu(x) \}$ of test functions carrying a Lorentz index, see \cite{Scharf2}
\begin{equation}
A(f) = \int d^4x\; A^{\mu}(x) f_\mu(x) \;, \label{smearA}
\end{equation}
where $\{ f_\mu(x) \}$ are required to transform in such a way to leave expression \eqref{smearA} Lorentz invariant. \\\\As it is apparent, the operator ${\cal A}_V$ displays the important property of being $BRST$ invariant, providing thus a  way to construct suitable $CHSH$ operators in order to investigate the violation of the $CHSH$ inequality in Higgs models within an explicit $BRST$ invariant environment. From the computational side, the operator ${\cal A}_V$ can be evaluated order by order in a loop expansion, much alike the usual way we deal with the perturbative treatment of the Wilson loop ${\cal W}_\gamma = e^{i \int_\gamma dx^\mu A_\mu}$. \\\\Let us end this section by giving a short  account of what we are currently doing on the $U(1)$ Higgs model,  whose detailed analysis will be reported in a forthcoming work \cite{prep}. \\\\As we have learned from the pioneering work \cite{Summers:1987fn,Summ,Summers:1987ze,Summers:1988ux,Summers:1995mv}, free fields are already able to produce a violation of the $CHSH$ inequality. Therefore, as a first step, we are looking at the purely quadratic part of the Higgs action $S_{\textrm{Higgs}} $, eq.\eqref{higgs_action}, namely 
\begin{eqnarray}
S_{\textrm{Higgs}}^{\textrm{quad}} & = & \int d^{4}x\left[-\frac{1}{4}F_{\mu\nu}(A)F^{\mu\nu}(A)+\frac{m^{2}}{2}A_{\mu}A^{\mu}+\frac{1}{2}\partial_{\mu}h\partial^{\mu}h-\frac{m_{h}^{2}}{2}h^{2} \right. \nonumber \\ 
& & \left. +\frac{1}{2}\partial_{\mu}\rho\partial^{\mu}\rho+mA_{\mu}\partial^{\mu}\rho+b\partial_{\mu}A^{\mu}-\overline{c}\partial^{2}c\right] \;, \label{Higg_quad}
\end{eqnarray}
where $m^2= e^2v^2$ and $m^2_h= \lambda v^2$ are the masses of the gauge vector boson and of the Higgs field $h$. \\\\Even at the quadratic level, the action $S_{\textrm{Higgs}}^{\textrm{quad}}$ exhibits an exact $BRST$ invariance, corresponding to the linear part of the transformations of eqs.\eqref{BRST_transf}, {\it i.e.} 
\begin{equation} 
s_{0} S_{\textrm{Higgs}}^{\textrm{quad}} = 0 \;, \qquad s_{0}s_{0}=0  \;, \label{soinv}
\end{equation} 
where 
\begin{eqnarray}
s_{0}A_{\mu} & = & -\partial_{\mu}c\,,\nonumber \\
s_{0}h & = & 0\,,\nonumber \\
s_{0}\rho & = & evc\,,\nonumber \\
s_{0}c & = & 0\,,\nonumber \\
s_{0}\overline{c} & = & b\,,\nonumber \\
s_{0}b & = & 0\,. \label{BRST_linear}
\end{eqnarray}
At the same order, the vector operator $V_\mu$, eq.\eqref{V_op}, becomes 
\begin{equation}
V^{lin}_{\mu}=\frac{1}{2}v\left(  \partial_{\mu} \rho + e v A_{\mu} \right) \label{V_linear} \;, 
\end{equation}
with 
\begin{equation} 
s_{0} V^{lin}_{\mu} = 0  \;. \label{invvo} 
\end{equation}
One easily recognizes that $V^{lin}_{\mu}$ coincides precisely with the physical part of the gauge boson field, displaying the content of the Higgs mechanism: the Goldstone mode is eaten by the gauge field, which becomes massive. \\\\We underline that the quadratic action in eq.\eqref{Higg_quad} can be canonically quantized by following the well known Kugo-Ojima procedure \cite{Kugo:1977mm,Kugo:1977mk,Kugo:1977yx,Kugo:1979gm} for  the construction of the physical Fock space with positive norm states through the extensive use of the cohomolgy of the $BRST$ charge.  We can therefore repeat the same analysis done in the previous sections and built a $s_{0}$-invariant  $CHSH$ correlator  by means of the $s_{0}$-invariant Weyl operator 




\begin{eqnarray}
{\cal A}_{V^{lin} } =  e^{ i {\hat V}^{lin}(f) } \label{Weyl_Vvo} \;, \qquad    s_{0}{\cal A}_{V^{lin} }=0 \;,    \label{weylvo}
\end{eqnarray}
allowing thus to investigate the possible violation of the $CHSH$ inequality in the $U(1)$ Higgs system already at the quadratic level  \cite{prep}. \\\\We mention here that a $BRST$ invariant operator $O$ can be introduced also for the Higgs field $h$ \cite{Dudal:2021dec,Dudal:2021pvw,Dudal:2020uwb,Capri:2020ppe,Dudal:2019pyg,Dudal:2019aew}
\begin{equation} 
O(x) = \frac{1}{2} \left( 2 v h(x) + h^2(x) + \rho^2(x) \right) \;, \qquad sO=0 \;. \label{Oh}
\end{equation} 
The operator $O$ shares many of the properties of the operator $V_\mu$ at the quantum level \cite{Dudal:2021dec,Dudal:2021pvw,Dudal:2020uwb,Capri:2020ppe,Dudal:2019pyg,Dudal:2019aew} and can be employed in order to have a $BRST$ invariant description of the Higgs field $h$. As in the case of $V_\mu$, $BRST$ invariant Weyl operators can be introduced by exponentiang $O$: 
\begin{equation} 
{\cal A}_O  =  e^{i {\hat O(g)}}   = e^{i \int_\Omega d^4x \; g(x) {\hat O(x) }} \label{Weyl_O}
\end{equation} 
where ${\hat O}$ denotes the dimensionless quantity 
\begin{equation} 
{\hat O}(x) = \frac{1}{v^2} O(x) \;. \label{hato}
\end{equation} 
Therefore, even in the quadratic approximation, it will be possible to investigate the violation of the $CHSH$ inequality by using Weyl operators of the Higgs type, eq.\eqref{hato}.

\section{Conclusion}
In this work we have analysed the violation of the $CHSH$ inequality in a relativistic Quantum Field Theory model. Following the pioneering work of \cite{Summers:1987fn,Summ,Summers:1987ze,Summers:1988ux,Summers:1995mv}, we started with
a pair of free massive real scalar fields $(\varphi_A^i, \varphi_B^i)$, $i=1,2,3$, taken  in the adjoint representation of the $SU(2)$ group, eq.\eqref{clact}. These fields have been employed to introduce a $CHSH$ type operator ${\cal C}^{aa'bb'}_{(ff'gg')} $, eqs.\eqref{Uab},\eqref{HU},\eqref{UUf}, obtained by means of   Hermitian combinations of Weyl operators. Making use of the canonical quantization, the correlation function  $\langle {\cal C}^{aa'bb'}_{(ff'gg')} \rangle$ of the above mentioned operator has been evaluated in closed form, eq.\eqref{chsh_exact}, allowing us to already detect a violation of the $CHSH$ inequality in the free case, see 
Fig.\eqref{violation_figure1} and Fig.\eqref{violation_figure2}. \\\\Although the reported violation turns out to be rather small as compared to Tsirelson's bound, we believe that the present work might be helpful for the investigation of more physical models. \\\\In particular, as discussed in section \eqref{Hg}, we have paid attention to devise a $BRST$ invariant framework for the study of the violation of the $CHSH$ inequality in the case of gauge theories, taking as explicit example the $U(1)$ Higgs model \cite{prep}. We highlight that the setup presented in section \eqref{Hg} generalizes as well to the case of the non-Abelian $SU(2)$ model \cite{Dudal:2021dec,Dudal:2021pvw,Dudal:2020uwb,Capri:2020ppe,Dudal:2019pyg,Dudal:2019aew}, a feature which might lead to a Quantum Field Theory investigation of the $CHSH$ inequality in the electroweak theory, a subject of great phenomenological and experimental interest, see \cite{Ashby-Pickering:2022umy} and refs. therein. \\\\A second topic which we are starting to look at is the possibility of obtaining a formualtion of the $CHSH$ inequality by means of direct use of the Feynman path integral. This would enable us to treat interacting field theories through the usual dictionary of the Feynman diagrams. Even if the task might seem to not present much difficulties, it requires, nevertheless, to face the challenging  issue of the renowed  lack of causality of the Feynman propagator $\Delta_F(x-y)$ \cite{Itzykson:1980rh}, namely 
\begin{equation}
\Delta_F(x-y)  = 
     \begin{cases}
     \frac{1}{4\pi} \delta((x-y)^2) -\frac{m}{8\pi \sqrt{(x-y)^2}} H^{(2)}_1\left (m\sqrt{(x-y)^2}\right)  \;, \qquad (x-y)^2 \ge 0 \\
       \frac{im}{4\pi^2 \sqrt{(-(x-y)^2)}} K_1\left( m \sqrt{-(x-y)^2} \right)\;,   \qquad (x-y)^2 < 0 \;,  \label{Feyn}
     \end{cases} 
    \end{equation}
where $H^{(2)}_1$ is the Hankel function, while $K_1$ is the modified Bessel function.  Expression \eqref{Feyn} shows the lack of causality of $\Delta_F(x-y)$: it receives non-vanishing contributions from the space-like region $(x-y)^2 <0$. This feature requires a deeper understanding of the relationship between entanglement and Feynman propagator, see for instance the discussion of ref.\cite{Fran}. This is certainly a topic worth to be investigated, due to the large amount of applications of the Feynman path integral in Quantum Field Theory. \\\\Finally, we would like to add to this short list the establishment of a Quantum Field Theory version of Bell's inequality \cite{Bell:1964kc}, which we reproduce below in its original form \cite{Bell:1964kc}:
\begin{equation} 
\large| E(a,b) - E(a,c) \large| \le 1 + E(b,c) \;, \label{Bi}
\end{equation}
where $E(a,b)$ stands for the expectation value of the product of the Alice and Bob measurements \cite{Bell:1964kc}. From the work of \cite{Summers:1987fn,Summ,Summers:1987ze,Summers:1988ux,Summers:1995mv}, we have learned how to formulate the $CHSH$ inequality in relativistic Quantum Field Theory. Though, we are unaware of a similar formulation for the original Bell inequality. This would be a nice achievement, in view of the pivotal role played by this inequality in the physics of entanglement. 

\section*{Acknowledgements}
The authors would like to thank the Brazilian agencies CNPq and FAPERJ for financial support. This study was financed in part
by the Coordena{\c c}{\~a}o de Aperfei{\c c}oamento de Pessoal de N{\'\i}vel Superior--Brasil (CAPES) --Finance Code 001. S.P.~Sorella is a level $1$ CNPq researcher under the contract 301030/2019-7. This work was performed during the recovery of one of the authors, S.P. Sorella, from a delicate eye surgery. He is particularly grateful to Alexa, Siri and Google assistant which have much facilitated his life during the post-surgery months.

\end{document}